\documentclass[letterpaper,10pt]{article} 

\usepackage{osameet3} %

\usepackage{amsmath}
\usepackage{tikz,pgfplots}
\usepackage[font={footnotesize}]{caption}
\usepackage{graphicx}
\usepackage{placeins}
\usepackage{mathtools}
\usepackage{wrapfig}
\usepackage{titlesec}
\usetikzlibrary{backgrounds,shapes,arrows,positioning}
\usetikzlibrary{shapes,snakes}
\usetikzlibrary{fit,calc} %

\usepackage[colorlinks=true,bookmarks=false,citecolor=blue,urlcolor=blue]{hyperref} %
\usepackage{textcomp}
\usepackage{acronym}
\let\mathcal\undefined
\DeclareMathAlphabet{\mathcal}{OMS}{cmsy}{m}{n}
\newcommand\blfootnote[1]{%
  \begingroup
  \renewcommand\thefootnote{}\footnote{#1}%
  \addtocounter{footnote}{-1}%
  \endgroup
}

\newcommand{\CW}{\mathcal{C}}

\newcommand{\dmin}{d_{\textnormal{min}}}
\newcommand{\ddesign}{d_{\textup{des}}}

\newcommand{\que}{\mathord{?}}

\newcommand{\ZO}{\{0, 1\}}
\newcommand{\ZQO}{\{0, \que, 1\}}

\usepackage{bm}
\renewcommand{\vec}[1]{\bm{#1}}

\newcommand{\BDD}{\textnormal{BDD}}

\newcommand{\dnE}[1]{\operatorname{d}_{\sim{\textnormal{E}(#1)}}}
\newcommand{\Sp}{\mathcal{S}}
\newcommand{\E}{\textnormal{E}}

\newcommand{\Eb}{E_{\textnormal{b}}}
\newcommand{\No}{N_{\textnormal{0}}}
\titlespacing*{\section}
{0pt}{0.3ex plus 0ex minus .2ex}{0.3ex plus 0ex minus .2ex}
\titlespacing*{\subsection}
{0pt}{0.3ex plus 0ex minus .2ex}{0.3ex plus 0ex minus .2ex}
\pgfplotsset{compat=1.17}
\definecolor{KITpalegreen}{RGB}{130,190,60}
\definecolor{KITcyanblue}{RGB}{80,170,230}
\definecolor{KITorange}{rgb}{.87,.60,.10}
\definecolor{mygreen}{rgb}{0.69, 0.87, 0.541}
\definecolor{myblue}{rgb}{0,0.4470,0.7410}
\definecolor{myblack}{rgb}{0.2,0.2,0.2}

\begin{document}

\acrodef{HDD}{hard-decision  decoding}
\acrodef{TPD}{turbo product decoding}
\acrodef{iBDD}{iterative bounded distance decoding}
\acrodef{BDD}{bounded distance decoding}
\acrodef{SA-HDD}{soft-aided \ac{HDD}}
\acrodef{PC}{product code}
\acrodef{AD}{anchor decoding}
\acrodef{HRB}{highly-reliable bit}
\acrodef{EaE}{error-and-erasure}
\acrodef{EaED}{\ac{EaE} decoder}
\acrodef{SABM}{soft-aided bit marking}
\acrodef{BI-AWGN}{binary-input additive white Gaussian noise}
\acrodef{BCH}{Bose--Chaudhuri--Hocquenghem}
\acrodef{DRS}{dynamic reliability score}
\acrodef{DRSD}{dynamic reliability score decoder}
\acrodef{BER}{bit error rate}
\acrodef{SABM-SR}{SABM with scaled reliabilities}
\acrodef{NCG}{net coding gain}

\title{Improved Soft-aided Error-and-erasure Decoding of Product Codes with Dynamic Reliability Scores}
    \vspace*{-3ex}
	\author{Sisi Miao, Lukas Rapp, and Laurent Schmalen}%
	\address{Karlsruhe Institute of Technology (KIT), Communications Engineering Lab (CEL), 76187 Karlsruhe, Germany}
	\email{\texttt{sisi.miao@kit.edu}}
	
	\copyrightyear{2022}
	
	\vspace*{-3ex}

\begin{abstract}
We propose a novel soft-aided low-complexity decoder for product codes based on dynamic reliability scores and error-and-erasure decoding. We observe coding gains of up to 1.2\,dB compared to conventional hard-decision decoders.\vspace*{-3ex}
\end{abstract}

\section{Introduction}
 \Acp{PC}~\cite{Elias1955} are powerful code constructions with high \acp{NCG} suitable for high-speed optical fiber communications. Soft-decision decoding of \acp{PC}, also known as \ac{TPD}~\cite{pyndiah1998near}, delivers excellent error-correcting performance at the cost of a very high internal decoder data flow with soft message-passing. In contrast, \ac{HDD}, in particular the ubiquitous \ac{iBDD} decoder, reduces the data flow significantly at the expense of a performance penalty. Recently, various hybrid algorithms have been proposed that use a certain amount of soft information to improve the coding gain of \ac{HDD} while keeping the decoder data flow manageable. 
A promising approach to improve the decoding performance of \acp{PC} is to use \ac{EaE} decoding, e.g.,~\cite{rapp2021error} and~\cite{soma2021errors}. A third channel output symbol, the ``erasure'', provides an elegant way to represent and update the bits with very low channel reliability. However, the additional coding gain is small, mostly because of the lack of miscorrection control.
The BEE-PC \cite{sheikh2021novel} decoder uses \ac{EaE} decoding with a relatively complex miscorrection control to improve the coding gain of \acp{PC}. A simpler approach to miscorrection-detection is to make use of the conflict between row/column decoding results and the bits that are considered very likely to be correct. These bits are called anchor bits in \ac{AD}~\cite{hager2018approaching} and \acp{HRB} in \ac{SABM} decoding~\cite{lei2019improved}. The latter is improved to \ac{SABM-SR} in~\cite{liga2019novel} for \acp{PC}. The difference between the two families of algorithms lies in the marking of these bits. For \ac{AD}, no channel reliability is used, the decoder sets all successfully decoded bits as anchor bits together with a back-tracking mechanism. 
In \ac{SABM}, the \acp{HRB} are set according to the channel reliability during the the initialization of decoding. Both algorithms provide good miscorrection-detection. Several other schemes have been proposed for PCs with varying degrees of performance-complexity trade-off~\cite{sheikh2021refined,sheikh2018low,sheikh2019binary}.

In this paper, we improve \ac{EaE} decoding from~\cite{rapp2021error} by introducing a simple and low-cost miscorrection control which resembles a combination of \ac{AD} and \ac{SABM}. We show coding gain improvements of $0.2$\,dB to SABM-SR decoding with significantly reduced complexity. \blfootnote{This work has received funding from the European Research Council (ERC)
under the European Union’s Horizon 2020 research and innovation programme
(grant agreement No. 101001899).}
\section{Preliminaries}
\label{sec:pre}
We consider \acp{PC} of rate $r = {k^2}/{n^2}$ whose each row/column vector $\vec{x}$ is a codeword of an $(n,k,t)$ component code $\mathcal{C}$, where $\mathcal{C}$ is either a $(2^{\nu}-1,k_0,t)$ binary \ac{BCH} code or its $(2^{\nu}-1,k_0-1,t)$ even-weight subcode able to correct $t$ errors. Let $\ddesign$ ($\ddesign \leq \dmin$) be the design distance of $\CW$ and $t=\lfloor (\ddesign-1)/2 \rfloor$. The codewords are transmitted over a \ac{BI-AWGN} channel which outputs $\tilde{y}_i=(-1)^{x_i}+n_i$, where $n_i$ is (real-valued) AWGN with noise variance $\sigma^2 = (2r\Eb/\No)^{-1}$. To obtain the discrete channel output $y_i\in\ZQO$, the values $\tilde{y}_i\in [-T,+T]$ are declared as erasures ``$\que$'', where $T$ is a configurable threshold. Values outside this interval are mapped to $0$ and $1$ by the usual \ac{HDD} rule.

We define
$
    \Sp^3_t(\vec{c}) \coloneqq  \{\vec{y} \in \ZQO^n : 2 \dnE{\vec{y}}(\vec{y}, \vec{c}) + \E(\vec{y}) < \ddesign\},
$
as the Hamming sphere in $\ZQO^n$ for a codeword $\vec{c} \in \CW$ where $\E(\vec{y}) :=|\{i : y_i = \que \}|$ is the number of erasures of $\vec{y}$ and $\dnE{\vec{y}}(\vec{y}, \vec{c})$ is the Hamming distance between $\vec{y}$ and $\vec{c}$ at the unerased coordinates of $\vec{y}$. We use the following \ac{EaED} which is a modification of~\cite[Sec. 3.8.1]{MoonBook}. Let $\vec{y} \in \ZQO^n$ be the received row/column vector and $\vec{w} \coloneqq  \textnormal{EaED}(\vec{y})$. If $\E(\vec{y})\geq \ddesign$, we do not decode and return $\vec{w}=\vec{y}$. If $\E(\vec{y})< \ddesign$, the erasure positions of $\vec{y}$ are first filled with two complementary random vectors in $\ZO^{\E(\vec{y})}$, resulting in two words $\vec{y}_1$,$\vec{y}_2 \in \ZO^n$. Note that the use of distinct random vectors is crucial for the performance of the decoder. Then, two \ac{BDD} steps are performed. Let $\vec{w}_1 \coloneqq \BDD(\vec{y}_1$) and $\vec{w}_2 \coloneqq \BDD (\vec{y}_2)$ and let $d_1=\dnE{\vec{y}}(\vec{y}, \vec{w}_1)$ and $d_2=\dnE{\vec{y}}(\vec{y}, \vec{w}_2)$. If both \ac{BDD} steps fail, set $\vec{w}=\vec{y}$. If $\vec{w}_i\in \CW$ for exactly one $\vec{w}_i$, set $\vec{w}=\vec{w}_i$. If both BDD succeed then $\vec{w}=\vec{w}_1$ if $d_1<d_2$ and $\vec{w}=\vec{w}_2$ if $d_1>d_2$;  If $d_1=d_2$, one of the codewords $\vec{w}_i$ is chosen at random. Such an \ac{EaED} is able to correct any joint \ac{EaE} pattern if $\vec{y}\in \Sp^3_t(\vec{c})$~\cite[Theorem 1]{rapp2021error}. Moreover, \ac{EaED} can correct some joint \ac{EaE} patterns for $2\dnE{\vec{y}}(\vec{y}, \vec{c}) + \E(\vec{y})\geq \ddesign$ because all (or most) of the erasures may possibly be filled with a correct value when generating $\vec{y}_1$ and $\vec{y_2}$. Thus, \ac{EaED} has potentially higher error-correcting capabilities than a one-step \ac{EaE} decoder~\cite{Forney1965} but is also more prone to miscorrections without the constraint that $\vec{w}\in \Sp^3_t(\vec{y})$. 
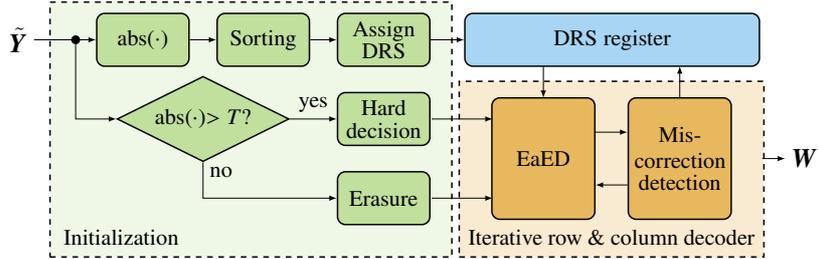
\begin{wrapfigure}{r}{0.7\textwidth}
        \centering
        \vspace{2ex}
     \scalebox{0.7}{\tikzset{block/.style={draw, thick, text width=1.5cm, minimum height=1cm, align=center, rounded corners=1.5mm,fill=white},
line/.style={-latex}   %
}  
\begin{tikzpicture}[tight background]

    \node[draw=none] (input) {\Large $\tilde{\vec{Y}}$};
    \node[draw,circle,fill,inner sep=1.5pt, right=0.7cm of input] (intersec) {};
    
    \node[block,right=0.3cm of intersec,fill=KITpalegreen!50] (abs) {\large abs($\cdot$)};  
    \node[block,right=0.5cm of abs,fill=KITpalegreen!50] (sort) {\large Sorting}; %
    \node[block,right=0.5cm of sort,fill=KITpalegreen!50] (assign) {\large Assign \\ DRS};  
    
    \node [draw, diamond, fill=white, thick, aspect=2,fill=KITpalegreen!50](condition)at ([yshift=-1.5cm]$(abs)!0.5!(sort)$) {\large abs($\cdot$)$>T$?};
    \node[block,fill=KITpalegreen!50] (hd)  at ([yshift=-1.5cm]$(sort)!1.0!(assign)$)  {\large Hard\\ decision}; 
    \node[block,fill=KITpalegreen!50] (erasure) at ([yshift=-1.5cm]$(abs)!1.0!(hd)$) {\large Erasure}; 
    \node[draw, thick, text width=1.7cm, minimum height=2.3cm, align=center, ,fill=KITorange!70, rounded corners=1.5mm] (eaed) at ([xshift=3cm]$(hd)!0.5!(erasure)$){\large EaED};  
    \node[draw, thick, text width=1.7cm, minimum height=2.3cm, align=center,right=0.6cm of eaed, fill=KITorange!70, rounded corners=1.5mm] (md) {\large Mis-\\correction\\ detection};  
    \node[block,draw, thick, text width=5.3cm, minimum height=1cm, align=center,fill=KITcyanblue!50, right=0.63cm of assign] (reg) {\large DRS register};  
    
    \node[draw=none, right=of md] (output) {\Large $\vec{W}$};

    \node[draw=none] at ([yshift=-1.5cm]$(eaed)!0.5!(md)$)(decodertext) {\large Iterative row \& column decoder};
    \node[draw=none, left=5.2cm of decodertext] {\large Initialization};
    
    \begin{scope}[on background layer]
    \node[draw,inner xsep=6mm,inner ysep=5mm,dashed,thick, fit=(eaed)(md),align=left,yshift=-2mm,fill=KITorange!20](backgroundDecoder){};
    \node[draw,fill=mygreen, fill opacity=0.2,inner xsep=4mm,inner ysep=4mm,dashed,thick,fit=(intersec)(abs)(sort)(assign)(condition)(hd)(erasure),align=left,yshift=-2mm](backgroundIni){};
    \end{scope}

    \draw[line] (abs)-- (sort);  
    \draw[line] (sort)-- (assign);  
    \draw[line] (assign) -- (reg); %
    \draw[line] (condition) --(hd) node[above,midway] {\large yes};
    \draw[line] (condition)  |- (erasure); 
    \node[draw=none, align=left]at ([yshift=-1cm]$(condition)!0.1!(hd)$) (no) {\large no};
    \draw[line] ($(hd)+(0.9,0)$)-- ($(eaed)+(-0.95,0.75)$);
    \draw[line] ($(erasure)+(0.9,0)$)-- ($(eaed)+(-0.95,-0.75)$);
    \draw[line] (intersec) |- (condition);
    \draw[line] (input)-- (abs);
    \draw[line] ($(md)+(1.55,0)$)-- (output);
    \draw[line] ($(eaed)+(0,1.75)$)-- ($(eaed)+(0,1.15)$);
    \draw[line] ($(md)+(0,1.15)$)-- ($(md)+(0,1.75)$);

    \draw  [line]([yshift=0.5cm]  eaed.east) -- ([yshift=0.5cm] md.west);
    \draw [line]([yshift=-0.5cm] md.west) -- ([yshift=-0.5cm] eaed.east);

\end{tikzpicture}  }  

    	\caption{Block diagram of the proposed DRSD}
    	\vspace{-1.1ex}
    	\label{fig:blockDiagram}
\end{wrapfigure} 
\vspace{-1.5ex}
\section{Proposed Algorithm}
We introduce a \emph{\ac{DRS}} for all the bits. The \acp{DRS} are stored in an additional register. The \ac{DRS} reflects
the reliability of a bit from both its channel output reliability and its behavior during the decoding. The \ac{DRS} is defined by an integer in the range $[0,31]$ such that it can be represented with $5$ bits. We manually set a threshold $T_{\mathrm{a}}$. Following~\cite{hager2018approaching}, all bits with a \ac{DRS}~$> T_{\mathrm{a}}$ are classified as anchor bits during decoding and are not allowed to be flipped by a component code decoder.

Figure \ref{fig:blockDiagram} depicts the block diagram and workflow of the proposed decoder. At the initialization, the received \ac{PC} word $\tilde{\vec{Y}}$ is fed to two paths. In the upper path, for all bits in  $\tilde{\vec{Y}}$, the absolute values $\{|\tilde{Y_i}|:i\in \{1,2,\ldots, n^2\}\}$ are sorted ascendingly and then evenly divided into $16$ groups (allowing the last group to have fewer entries than the others if $n^2$ does not divide 16). To each group, we assign a \ac{DRS} in the range of $[9,24]$ accordingly. The bits with lowest $|\tilde{Y}_i|$ will have \ac{DRS} $9$ while the bits with highest $|\tilde{Y}_i|$ will have a \ac{DRS} of $24$. This initial \ac{DRS} value is stored in the \ac{DRS} register. In the lower branch, erasures are marked if $|\tilde{Y}_i|<T$, with  $T$ (defined in Sec.~\ref{sec:pre}) an optimizable threshold. For non-erased bits, a usual hard-decision  is performed. The values in $\ZQO$ are passed to the iterative row and column decoder.

Iterative row and column decoding is performed with an EaED for the component codes. Additionally, after every decoding step, the miscorrection detection unit evaluates whether an anchor bit is flipped by the decoding decision. In this case, this decision is discarded and the \ac{DRS} for all the anchor bits in conflict is reduced by one. If a decoding step does not flip any anchor bit, this decision will be accepted while the \ac{DRS} of all flipped bits is reduced by one. If a vector is already a codeword and thus no decoding is performed, the \ac{DRS} of every bit in it is increased by one. 
In the case of a decoding failure, neither the codeword nor the \ac{DRS} is changed. In addition, the threshold $T_{\mathrm{a}}$ is increased by $1$ every five decoding iterations, such that a small penalty is given for words that fail to decode consistently. With the update of \acp{DRS}, the anchor bits are reevaluated every iteration. A simple yet elegant updating of both the unreliable (via erasures) and highly-reliable bits (via \acp{DRS}) is achieved. We call this new algorithm \ac{DRSD}.

During the decoding, only hard messages are passed. For \ac{EaED}, ternary messages are used. For the communication between the \ac{DRS} register and the decoder, binary messages are sent from the \ac{DRS} register to the \ac{EaED} representing whether a bit is an anchor or not and from the \ac{EaED} to the \ac{DRS} register representing an increase/decrease of the \ac{DRS}. However, the information flow is still higher than \ac{iBDD}.
The major computational overhead of our algorithm comparing to \ac{iBDD} is the usage of \ac{EaED}, where two \ac{BDD} steps are performed with the presence of erasures for every row/column decoding. This increases the total number of \acp{BDD}, especially in the first few iterations, before the erasures are resolved. Words without erasures can be decoded with conventional \ac{BDD}.
The additional storage for \acp{DRS} is relatively small as well, as the \acp{DRS} are stored with 5-bit integers.

\section{Simulation Results}
\label{sec:simulation results}
For simulation, we consider the two cases with $10$ or $20$ decoding iterations. For $10$ iterations, we decode with $8$ iterations of \ac{DRSD} first, followed by $2$ simple \ac{EaED} iterations not using the \ac{DRS} register; For $20$ iterations, we use $16$ iterations of \ac{DRSD} followed by $4$ plain \ac{EaED} iterations. This eliminates the influence of erroneous bits with high \ac{DRS}. We also simulate an \emph{ideal} EaED with a genie-aided miscorrection detection where miscorrections are always discarded to benchmark our results.

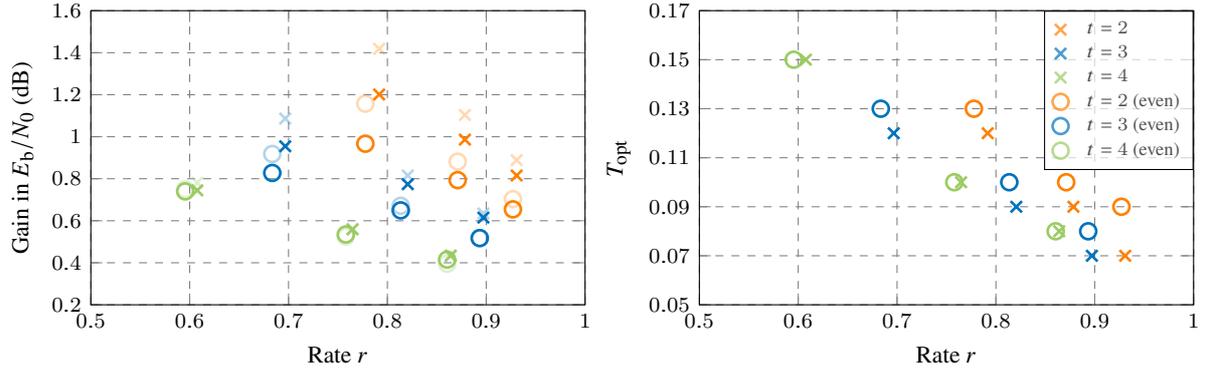
\begin{figure}[htbp]
        \centering
        \vspace{-1ex}
    	\tikzset{mark options={mark size=3, line width=1pt}}
\begin{tikzpicture}
\pgfplotsset{grid style={dashed, gray},scaled y ticks=false}
\pgfplotsset{every tick label/.append style={font=\footnotesize}}
\begin{axis}[%
        name=ax1,
        scatter/classes={%
        d={mark=x,draw=orange},e={mark=x,draw=myblue},f={mark=x,draw=KITpalegreen!80},a={mark=o,draw=orange},b={mark=o,draw=myblue},c={mark=o,draw=KITpalegreen!80}, dd={mark=x,draw=orange },ed={mark=x,draw=myblue },fd={mark=x,draw=KITpalegreen!80 },ad={mark=o,draw=orange },bd={mark=o,draw=myblue },cd={mark=o,draw=KITpalegreen!80}
        },
        xlabel={Rate $r$},
        ylabel={Gain in $\Eb/\No$ (dB)},
        scale only axis,
        xmin=0.5,
        xmax=1,
        ymin=0.2,
        ymax=1.6,
        width=6.5cm,
        height=3.9cm,
        yminorticks,
        axis background/.style={fill=white, mark size=2pt},
        xmajorgrids,
        ymajorgrids,
        yminorgrids,
        label style={font=\small},
        xtick={0.5,0.6,...,1},
        ytick={0.2,0.4,...,1.6},
        ]

\addplot[scatter,only marks,scatter src=explicit symbolic]table[meta=label] {
x y label
0.7777295554591109 0.9668000000000001 a
};
\addplot[scatter,only marks,scatter src=explicit symbolic]table[meta=label] {
x y label
0.6835513671027342 0.8276400000000002 b
};
\addplot[scatter,only marks,scatter src=explicit symbolic]table[meta=label] {
x y label
0.5954491908983818 0.7397455 c
};
\addplot[scatter,only marks,scatter src=explicit symbolic]table[meta=label] {
x y label
0.8711111111111111 0.7934600000000005 a
};
\addplot[scatter,only marks,scatter src=explicit symbolic]table[meta=label] {
x y label
0.8135332564398308 0.6494149999999999 b
};
\addplot[scatter,only marks,scatter src=explicit symbolic]table[meta=label] {
x y label
0.757923875432526 0.5346699999999993 c
};
\addplot[scatter,only marks,scatter src=explicit symbolic]table[meta=label] {
x y label
0.9270185086607359 0.6542999999999992 a
};
\addplot[scatter,only marks,scatter src=explicit symbolic]table[meta=label] {
x y label
0.8934133983861888 0.5175749999999999 b
};
\addplot[scatter,only marks,scatter src=explicit symbolic]table[meta=label] {
x y label
0.8604286901474796 0.4150350000000005 c
};
\addplot[scatter,only marks,scatter src=explicit symbolic]table[meta=label] {
x y label
0.7916795833591668 1.2011699999999994 d
};
\addplot[scatter,only marks,scatter src=explicit symbolic]table[meta=label] {
x y label
0.6966333932667866 0.9545900000000003 e
};
\addplot[scatter,only marks,scatter src=explicit symbolic]table[meta=label] {
x y label
0.6076632153264306 0.7446299999999999 f
};
\addplot[scatter,only marks,scatter src=explicit symbolic]table[meta=label] {
x y label
0.8784467512495194 0.9863250000000008 d
};
\addplot[scatter,only marks,scatter src=explicit symbolic]table[meta=label] {
x y label
0.8206228373702422 0.7739249999999998 e
};
\addplot[scatter,only marks,scatter src=explicit symbolic]table[meta=label] {
x y label
0.7647673971549405 0.5590799999999998 f
};
\addplot[scatter,only marks,scatter src=explicit symbolic]table[meta=label] {
x y label
0.9307907062243175 0.815430000000001 d
};
\addplot[scatter,only marks,scatter src=explicit symbolic]table[meta=label] {
x y label
0.8971166623902329 0.6152350000000002 e
};
\addplot[scatter,only marks,scatter src=explicit symbolic]table[meta=label] {
x y label
0.864063020591986 0.4345699999999999 f
};
\addplot[scatter,only marks,scatter src=explicit symbolic, draw opacity=0.3]table[meta=label] {
x y label
0.7777295554591109 1.157225  ad
};
\addplot[scatter,only marks,scatter src=explicit symbolic, draw opacity=0.3]table[meta=label] {
x y label
0.6835513671027342 0.9179700000000002 bd
};
\addplot[scatter,only marks,scatter src=explicit symbolic, draw opacity=0.3]table[meta=label] {
x y label
0.5954491908983818 0.749511 cd
};
\addplot[scatter,only marks,scatter src=explicit symbolic, draw opacity=0.3]table[meta=label] {
x y label
0.8711111111111111 0.8813500000000003  ad
};
\addplot[scatter,only marks,scatter src=explicit symbolic, draw opacity=0.3]table[meta=label] {
x y label
0.8135332564398308 0.6713899999999997 bd
};
\addplot[scatter,only marks,scatter src=explicit symbolic, draw opacity=0.3]table[meta=label] {
x y label
0.757923875432526 0.5249049999999995 cd
};
\addplot[scatter,only marks,scatter src=explicit symbolic, draw opacity=0.3]table[meta=label] {
x y label
0.9270185086607359 0.7031249999999991  ad
};
\addplot[scatter,only marks,scatter src=explicit symbolic, draw opacity=0.3]table[meta=label] {
x y label
0.8934133983861888 0.5175749999999999 bd
};
\addplot[scatter,only marks,scatter src=explicit symbolic, draw opacity=0.3]table[meta=label] {
x y label
0.8604286901474796 0.395505 cd
};
\addplot[scatter,only marks,scatter src=explicit symbolic, draw opacity=0.3]table[meta=label] {
x y label
0.7916795833591668 1.4184599999999996 dd
};
\addplot[scatter,only marks,scatter src=explicit symbolic, draw opacity=0.3]table[meta=label] {
x y label
0.6966333932667866 1.0864300000000002 ed
};
\addplot[scatter,only marks,scatter src=explicit symbolic, draw opacity=0.3]table[meta=label] {
x y label
0.6076632153264306 0.7812505000000001 fd
};
\addplot[scatter,only marks,scatter src=explicit symbolic, draw opacity=0.3]table[meta=label] {
x y label
0.8784467512495194 1.1035150000000007 dd
};
\addplot[scatter,only marks,scatter src=explicit symbolic, draw opacity=0.3]table[meta=label] {
x y label
0.8206228373702422 0.8154299999999992 ed
};
\addplot[scatter,only marks,scatter src=explicit symbolic, draw opacity=0.3]table[meta=label] {
x y label
0.7647673971549405 0.5615199999999998 fd
};
\addplot[scatter,only marks,scatter src=explicit symbolic, draw opacity=0.3]table[meta=label] {
x y label
0.9307907062243175 0.8886700000000012 dd
};
\addplot[scatter,only marks,scatter src=explicit symbolic, draw opacity=0.3]table[meta=label] {
x y label
0.8971166623902329 0.6347649999999998 ed
};
\addplot[scatter,only marks,scatter src=explicit symbolic, draw opacity=0.3]table[meta=label] {
x y label
0.864063020591986 0.4296850000000001 fd
};

\end{axis}

\begin{axis}[%
        at={(ax1.south east)},
        xshift=1.5cm,
        scatter/classes={%
    d={mark=x,draw=orange},e={mark=x,draw=myblue},f={mark=x,draw=KITpalegreen!80},a={mark=o,draw=orange},b={mark=o,draw=myblue},c={mark=o,draw=KITpalegreen!80}, dd={mark=x,draw=orange },ed={mark=x,draw=myblue },fd={mark=x,draw=KITpalegreen!80 },ad={mark=o,draw=orange },bd={mark=o,draw=myblue },cd={mark=o,draw=KITpalegreen!80 }
    },
        xlabel={Rate $r$},
        ylabel={$T_{\textnormal{opt}}$},
        scale only axis,
        width=6.5cm,
        height=3.9cm,
        xmin=0.5,
        xmax=1,
        ymin=0.05,
        ymax=0.17,
        yminorticks,
        axis background/.style={fill=white, mark size=1pt},
        xmajorgrids,
        ymajorgrids,
        yminorgrids,
        xtick={0.5,0.6,...,1},
        ytick={0.05,0.07,...,0.17},
        label style={font=\small},
        legend cell align={left},
        yticklabel style={
        /pgf/number format/fixed,
        /pgf/number format/precision=2},
         legend cell align={left},
        legend pos=north east,
        legend style={at = {(1,1)}, fill opacity=0.7, text opacity = 1,legend columns=1, row sep = -1pt, font=\scriptsize}]
\addlegendentry{$t=2\;\;$}
\addplot[scatter,only marks,scatter src=explicit symbolic]table[meta=label] {
x y label
0.6966333932667866 0.12 e
};
\addlegendentry{$t=3\;\;$}
\addplot[scatter,only marks,scatter src=explicit symbolic]table[meta=label] {
x y label
0.6076632153264306 0.15 f
};
\addlegendentry{$t=4\;\;$}
\addplot[scatter,only marks,scatter src=explicit symbolic]table[meta=label] {
x y label
0.8784467512495194 0.09 d
};
\addplot[scatter,only marks,scatter src=explicit symbolic]table[meta=label] {
x y label
0.8206228373702422 0.09 e
};
\addplot[scatter,only marks,scatter src=explicit symbolic]table[meta=label] {
x y label
0.7647673971549405 0.1 f
};
\addplot[scatter,only marks,scatter src=explicit symbolic]table[meta=label] {
x y label
0.9307907062243175 0.07 d
};
\addplot[scatter,only marks,scatter src=explicit symbolic]table[meta=label] {
x y label
0.8971166623902329 0.07 e
};
\addplot[scatter,only marks,scatter src=explicit symbolic]table[meta=label] {
x y label
0.864063020591986 0.08 f
};
\addplot[scatter,only marks,scatter src=explicit symbolic]table[meta=label] {
x y label
0.7777295554591109 0.13 a
};
\addlegendentry{$t=2$ (even)}
\addplot[scatter,only marks,scatter src=explicit symbolic]table[meta=label] {
x y label
0.6835513671027342 0.13 b
};
\addlegendentry{$t=3$ (even)}
\addplot[scatter,only marks,scatter src=explicit symbolic]table[meta=label] {
x y label
0.5954491908983818 0.15 c
};
\addlegendentry{$t=4$ (even)}
\addplot[scatter,only marks,scatter src=explicit symbolic]table[meta=label] {
x y label
0.8711111111111111 0.1 a
};
\addplot[scatter,only marks,scatter src=explicit symbolic]table[meta=label] {
x y label
0.8135332564398308 0.1 b
};
\addplot[scatter,only marks,scatter src=explicit symbolic]table[meta=label] {
x y label
0.757923875432526 0.1 c
};
\addplot[scatter,only marks,scatter src=explicit symbolic]table[meta=label] {
x y label
0.9270185086607359 0.09 a
};
\addplot[scatter,only marks,scatter src=explicit symbolic]table[meta=label] {
x y label
0.8934133983861888 0.08 b
};
\addplot[scatter,only marks,scatter src=explicit symbolic]table[meta=label] {
x y label
0.8604286901474796 0.08 c
};
\addplot[scatter,only marks,scatter src=explicit symbolic]table[meta=label] {
x y label
0.7916795833591668 0.12 d
};
\end{axis}
\end{tikzpicture}
    	\vspace{-0.1ex}
    	\caption{Simulation results of the parameter analysis: The  noise threshold gain that the \ac{DRSD} achieves compared to \ac{iBDD} is plotted on the left and the corresponding optimal threshold $T_{\textnormal{opt}}$ for erasures on the right. The component codes are $(n,k,t)-$\ac{BCH} code with $n\in \{127,255,511\}$ and $t\in \{2,3,4\}$ or their even-weight subcodes. The transparent markers corresponds to the respective ideal EaED.}
	    \label{fig:gain}
\end{figure}

\begin{figure}[htbp]
        \centering
        \vspace{1ex}
    	    \begin{tikzpicture}
       \pgfplotsset{grid style={dashed, gray}}
       \pgfplotsset{every tick label/.append style={font=\footnotesize}}
        \begin{axis}[%
        name=ax1,
        width=8cm,
        height=5.9cm,
        xmin=2.9,
        xmax=4.9,
        ymode=log,
        ymin=1e-7,
        ymax=0.1,
        yminorticks,
        axis background/.style={fill=white, mark size=1.5pt},
        xmajorgrids,
        ymajorgrids,
        yminorgrids,
        xtick={2.9,3.1,...,5.3},
        ytick={0.1,0.01,0.001,1e-4,1e-5,1e-6,1e-7,1e-8},
        ylabel={Post-FEC BER},
        xlabel={$\Eb/\No$ (dB)},
        label style={font=\small},
        legend cell align={left},
        legend style={anchor = north east, draw=none, fill opacity=0.7, text opacity = 1,legend columns=1, row sep = 0pt,font=\footnotesize}
]
 \addplot [color=orange, line width=0.9pt, mark=*, mark options={solid, orange, fill=white, mark size=1.5pt}]
   table[row sep=crcr]{%
3.03649 0.0351582\\
3.11983 0.0291384\\
3.20316 0.0213643\\
3.28649 0.0135962\\
3.36983 0.0069004\\
3.45316 0.00249271\\
3.53649 0.000454642\\
3.61983 5.12845e-05\\
3.70316 3.67004e-06\\
3.78649 2.16932e-07\\
3.86983 2.13473e-08\\
 };
 \addlegendentry{DRSD (10 it.)}
 
   \addplot [color=orange, line width=0.9pt, dashed, mark=*, mark options={solid, orange, fill=white, mark size=1.5pt}]table[row sep=crcr]{%
3.03649 0.0296026\\
3.11983 0.022856\\
3.20316 0.0141033\\
3.28649 0.00503048\\
3.36983 0.00150155\\
3.45316 0.000165315\\
3.53649 2.1766e-05\\
3.61983 1.02171e-06\\
3.70316 6.83611e-08\\
 };
\addlegendentry{DRSD (20 it.)}

 \addplot [color=myblack, line width=0.9pt, dotted, mark=square*, mark options={solid, myblack, fill=white, mark size=1pt}]
   table[row sep=crcr]{%
3.08565 0.015612\\
3.16898 0.00831108\\
3.22741 0.00408426\\
3.31074 0.00116427\\
3.39408 0.000150509\\
3.47741 1.07446e-05\\
3.56074 3.54967e-07\\
3.64408 8.95873e-09\\
};
\addlegendentry{ideal EaED}
\addplot [color=KITpalegreen!80, line width=0.9pt, mark=diamond*, mark options={solid,KITpalegreen!80,fill=white, mark size=1.5pt}]
  table[row sep=crcr]{%
3.44693 0.0322138\\
3.6136 0.0262655\\
3.78027 0.0220589\\
3.94693 0.015239\\
4.1136 0.00810881\\
4.28027 0.0016166\\
4.44693 0.000137922\\
4.6136 3.80668e-06\\
4.78027 1.04924e-07\\
4.93785 2.01883e-08\\
} node [pos=0.5,anchor=south,font=\footnotesize,sloped] {iBDD};

\addplot [color=myblue, line width=0.9pt, mark=triangle*, mark options={solid, myblue, fill=white, mark size=1.5pt}]
  table[row sep=crcr]{%
3.08263052328897	0.0435844623698019\\
3.18263052328897	0.0389537160310126\\
3.28263052328897	0.0330722844388754\\
3.38263052328897	0.0224191401049417\\
3.48263052328897	0.0139360952306367\\
3.58263052328897	0.0047877672488057\\
3.68263052328897	0.00121058814315921\\
3.78263052328897	8.37610975444792e-05\\
3.88263052328897	5.20241441221941e-06\\
3.98263052328897 1.86711340176776e-07\\
4.098263052328897 4.86711340176776e-09\\
} node [pos=0.5,anchor=south,font=\footnotesize,sloped] {SABM-SR};

\addplot[color=black, line width=0.9pt, mark=square*, mark options={solid, black, fill=white, mark size=1pt}]
  table[row sep=crcr]{%
2.9	0.0202655407357798\\
3	0.00509089398311894\\
3.1	0.000466080054785853\\
3.2	1.29053451418654e-05\\
3.3	4.71449588895136e-07\\
3.4	3.91569425992638e-09\\
} node [pos=0.5,anchor=south,font=\footnotesize,sloped] {TPD};
\draw[latex-latex,very thick] (axis cs:3.80,2.2e-7)--(axis cs:4.76, 2.2e-7);
\node[font=\footnotesize] at (axis cs:4.3,4e-7) {0.96\,dB};

\end{axis}

       \begin{axis}[%
       at={(ax1.south east)},
        xshift=1.5cm,
        xmin=3.5,
        xmax=5.3,
        ymode=log,
        ymin=1e-7,
        ymax=0.1,
        yminorticks=true,
        axis background/.style={fill=white, mark size=1.5pt},
        xmajorgrids,
        ymajorgrids,
        yminorgrids,
        width=8cm,
        height=5.9cm,
        xtick={3.5,3.7,...,5.5},
        ytick={0.1,0.01,0.001,1e-4,1e-5,1e-6,1e-7,1e-8},
        xlabel={$\Eb/\No$ (dB)},
        ylabel={Post-FEC BER},
        label style={font=\small},
        legend cell align={left},
        legend style={anchor = north east, draw=none, fill opacity=0.7, text opacity = 1,legend columns=1,font=\footnotesize, row sep = 0pt}
]
 \addplot [color=orange, line width=0.9pt, mark=*, mark options={solid, orange, fill=white, mark size=1.5pt}]
 table[row sep=crcr]{%
 	3.89162 0.0175631\\
 	3.97079 0.0135286\\
 	4.04995 0.00847037\\
 	4.12912 0.00311496\\
 	4.20829 0.00036694\\
 	4.28745 8.50604e-06\\
 	4.36662 3.0055e-08\\
 };
\addlegendentry{DRSD (10 it.)}

 \addplot [color=orange, dashed, line width=0.9pt, mark=*, mark options={solid, orange, fill=white, mark size=1.5pt}]
   table[row sep=crcr]{%
3.89162 0.0150249\\
3.97079 0.0089769\\
4.04995 0.00249982\\
4.12912 0.000224489\\
4.20829 3.30875e-06\\
4.28745 9.83661e-09\\
 };
 \addlegendentry{DRSD (20 it.)}

\addplot [color=myblack, line width=0.9pt, dotted, mark=square*, mark options={solid, myblack, fill=white, mark size=1pt}]
table[row sep=crcr]{%
	3.74986 0.016249\\
	3.82903 0.0136476\\
    3.89162 0.0109852\\
    3.97079 0.00663557\\
    4.04995 0.00201227\\
    4.12912 0.000242634\\
    4.20829 3.7249e-06\\
    4.28745 1.71726e-08\\
};
\addlegendentry{ideal EaED}

\addplot [color=KITpalegreen!80, line width=0.9pt, mark=diamond*, mark options={solid,KITpalegreen!80,fill=white, mark size=1.5pt}]
  table[row sep=crcr]{%
4.18365 0.0168525\\
4.34198 0.0140363\\
4.50032 0.0107882\\
4.65865 0.00708795\\
4.73782 0.00423753\\
4.81698 0.00200887\\
4.89615 0.000396122\\
5.01873 7.74358e-06\\
5.09789 2.14556e-07\\
5.17706 7.43846e-09\\
} node [pos=0.5,anchor=south,font=\footnotesize,sloped] {iBDD};;

\addplot [color=myblue, line width=0.9pt, mark=triangle*, mark options={solid, myblue, fill=white, mark size=1.5pt}]
  table[row sep=crcr]{
3.39684128727424	0.0351557220636894\\
3.49684128727424	0.032829957458728\\
3.59684128727424	0.0312758880271704\\
3.69684128727424	0.0289893384219464\\
3.79684128727424	0.0263360235289998\\
3.89684128727424	0.0237651651756797\\
3.99684128727424	0.0194737487088811\\
4.09684128727424	0.0153251868839831\\
4.19684128727424	0.00842142119360655\\
4.29684128727424	0.00157385199838938\\
4.39684128727424	5.06474029346528e-05\\
4.49684128727424	1.78271286267877e-07\\
4.59684128727424	6.78271286267877e-10\\
} node [pos=0.5,anchor=south,font=\footnotesize,sloped] {SABM-SR};;

\addplot[color=black, line width=0.9pt, mark=square*, mark options={solid, black, fill=white, mark size=1pt}]
table[row sep=crcr]{%
		3.2	0.054437587082942\\
		3.3	0.046984287279358\\
		3.4	0.042677157346018\\
		3.5	0.035219971639152\\
		3.6	0.030124939456009\\
		3.7	0.0213965988332\\
		3.8	0.00829490177189\\
		3.9	0.000259890294429629\\
		3.925 7.4173e-05\\
		3.955 1.3443e-05\\
		3.9745 2.4169e-06\\
		4     1.6083e-07\\	
		4.02     1e-08\\	
	} node [pos=0.5,anchor=south,font=\footnotesize,sloped] {TPD};;
\draw[very thick,latex-latex] (axis cs:4.33,2.2e-7)--(axis cs:5.10, 2.2e-7);
\node[font=\footnotesize] at (axis cs:4.8,4e-7) {0.76\,dB};

\end{axis}
\end{tikzpicture}  
    	\vspace{-2ex}
    	\caption{BER vs. $\Eb/\No$ (in dB) curve for product codes of rate $0.78$ (28\% overhead, left plot) and $0.87$ (15\% overhead, right plot).}
    	\vspace{-2ex}
	    \label{fig:BERcurve}
\end{figure}
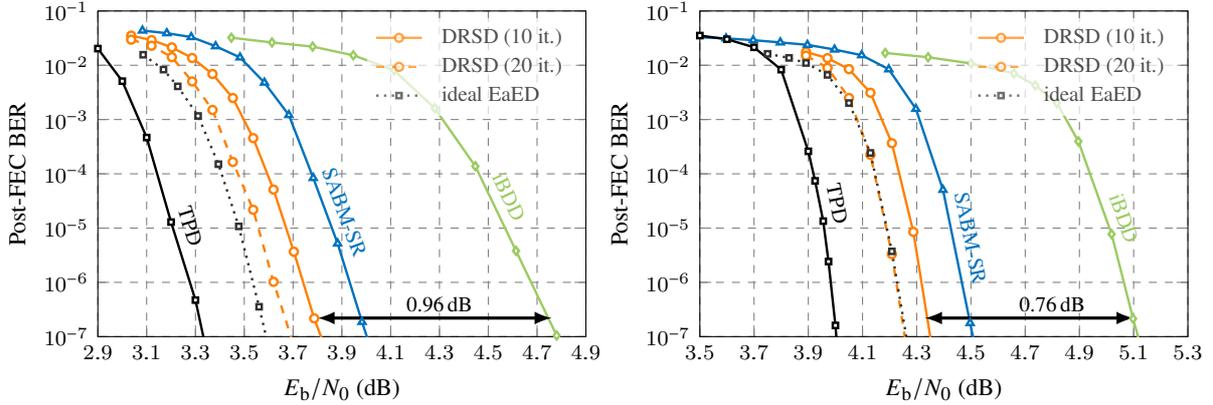
In Fig. \ref{fig:gain}, the noise threshold gain of \ac{DRSD} (opaque markers) for a target \ac{BER} at $10^{-4}$ over \ac{iBDD} is shown for various component codes as well as the respectively optimal erasure thresholds $T$ ($T_{\textnormal{opt}}$). We allow $20$ iterations for both decoders. The noise thresholds are estimated by a Monte Carlo method along with a binary search. We observe that slightly varying $T$ (within $\pm 10\%$) does not degrade the performance severely. The initial value of the anchor threshold $T_{\mathrm{a}}$ is set to be 9 for $t=2$, 10 for $t=3$, and 12 for $t=4$, except for the $(127,4)$ component codes where $T_{\mathrm{a}}=14$. The transparent markers stand for ideal EaED with 16 iterations (the optimal threshold $T$ may differ but is not plotted for the sake of clarity). One can see that the larger the $t$ and the higher the code rate, the closer the \ac{DRSD} performs to an ideal \ac{EaED}, which means our decoder is nearly miscorrection-free. Unfortunately, both decoders show smaller gains in this region.

In Fig. \ref{fig:BERcurve}, we compare the residual post-FEC \ac{BER} after 10 decoding iterations for different decoders (we additionally show the \ac{DRSD} performance after 20 iterations). We use even-weight subcodes of BCH codes as component codes denoted by $\mathcal{C}_1(127,112,2)$ and $\mathcal{C}_2 (255,238,2)$ with \ac{PC} rates of $0.78$ ($28$\% overhead) and $0.87$ ($15$\% overhead), respectively. For reference, we show the results of \ac{TPD} and \ac{SABM-SR} with the data points provided in ~\cite{liga2019novel} where the component codes are singly-extended BCH codes with parameters $(128,113,2)$ and $(256,239,2)$ (hence with a small, negligible rate difference compared to our codes). For both \acp{PC}, an obvious gain compared to \ac{iBDD} can be observed and the performance is further improved by running $20$ iterations. With twice the iterations, our decoder behaves very closely to an ideal \ac{EaED} and has only a small gap to the significantly more complex TPD. For the rate $0.78$ \ac{PC}, we estimate at a residual \ac{BER} of $10^{-15}$ a \ac{NCG} of $10.88$ dB ($10$ iterations) and $11.03$ dB ($20$ iterations). For the rate $0.87$ \ac{PC}, the conjectured \acp{NCG} are $10.51$\,dB and $10.59$\,dB respectively.

\section{Conclusions}
The proposed \ac{DRS} decoder outperforms other soft-aided \ac{HDD} schemes with a near miscorrection-free \ac{EaE} decoding keeping the complexity similarly low as conventional \ac{iBDD}. The \acp{NCG} make this scheme a promising candidate for future low-complexity optical communication systems. Future research directions include the extension of this scheme to staircase codes~\cite{Smith2012}. %

\end{document}